\documentclass[fleqn,usenatbib]{mnras}

\usepackage{mathptmx}

\usepackage[T1]{fontenc}

\DeclareRobustCommand{\VAN}[3]{#2}
\let\VANthebibliography\thebibliography
\def\thebibliography{\DeclareRobustCommand{\VAN}[3]{##3}\VANthebibliography}

\usepackage{graphicx}	
\usepackage{amsmath}	
\usepackage{amssymb}	
\usepackage{xcolor}

\definecolor{ochre}{rgb}{0.8, 0.47, 0.13}

\newcommand{\event}{TIC 470710327}

\newcommand{\ba}{\begin{eqnarray}}
\newcommand{\ea}{\end{eqnarray}}
\newcommand{\be}{\begin{equation}}
\newcommand{\ee}{\end{equation}}
\newcommand{\gr}{\mathrm{GR}}

\newcommand{\OUT}{\mathrm{out}}

\newcommand{\zlk}{\mathrm{ZLK}}
\newcommand{\A}{\mathrm{A}}
\newcommand{\B}{\mathrm{B}}

\newcommand{\li}{\mathrm{lim}}

\newcommand{\tide}{\mathrm{Tide}}
\def\e1{e_1^2}

\title[Stellar reduction for \event]{Mergers prompted by dynamics in compact, multiple-star systems: a stellar-reduction case for the massive triple \event}

\author[Vigna-G\'omez et al.]{
Alejandro Vigna-G\'omez,$^{1}$\thanks{E-mail: avignagomez@nbi.ku.dk}
Bin Liu,$^{1}$
David R. Aguilera-Dena,$^{2}$
Evgeni Grishin,$^{3,4}$
\newauthor
Enrico Ramirez-Ruiz$^{5}$ and
Melinda Soares-Furtado$^{6}$
\\
$^{1}$Niels Bohr International Academy, 
Niels Bohr Institute, 
Blegdamsvej 17, DK-2100, 
Copenhagen, Denmark\\
$^{2}$Institute of Astrophysics, 
FORTH, Dept. of Physics, 
University of Crete, 
Voutes, University Campus, GR-71003,
Heraklion, Greece\\
$^{3}$School of Physics and Astronomy, Monash University, Clayton, VIC 3800, Australia\\
$^{4}$The ARC Centre of Excellence for Gravitational Wave Discovery -- OzGrav, Clayton, VIC 3800, Australia\\
$^{5}$Department of Astronomy and Astrophysics,
University of California,
Santa Cruz, CA 95064, USA\\
$^{6}$Department of Astronomy, 
University of Wisconsin-Madison, 
475 N. Charter St., Madison, 
WI 53703, USA
}

\begin{document}
\label{firstpage}
\pagerange{\pageref{firstpage}--\pageref{lastpage}}
\maketitle

\begin{abstract}
\event, a massive compact hierarchical triple-star system, was recently identified by NASA's Transiting Exoplanet Survey Satellite (TESS).
\event\ is comprised of a compact (1.10 d) circular eclipsing binary, with total mass $\approx 10.9-13.2\ \rm{M_{\odot}}$, and a more massive ($\approx 14-17\ \rm{M_{\odot}}$) eccentric non-eclipsing tertiary in a 52.04 d orbit.
Here we present a progenitor scenario for \event\ in which `2+2' quadruple dynamics result in Zeipel-Lidov-Kozai (ZLK) oscillations that lead to a contact phase of the more massive binary.
In this scenario, the two binary systems should form in a very similar manner, and dynamics trigger the merger of the more massive binary either during late phases of star formation or several Myr after the zero-age main sequence (ZAMS), when the stars begin to expand.
Any evidence that the tertiary is a highly-magnetised ($\sim 1-10$ kG), slowly-rotating blue main-sequence star would hint towards a quadruple origin.
Finally, our scenario suggests that the population of inclined, compact multiple-stellar systems is reduced into co-planar systems, via mergers, late during star formation or early in the main sequence.
The elucidation of the origin of \event\ is crucial in our understanding of multiple massive-star formation and evolution.
\end{abstract}




\section{Introduction}
\cite{2022MNRAS.511.4710E} reported the discovery of \event, initially identified in data obtained TESS and followed up with the HERMES spectrograph.
\event\ is comprised of a 1.10~d circular binary and a massive tertiary on a $52.04$~d orbit.
The close binary is constituted of main-sequence stars with estimated individual masses of $6-7\ \rm{M_{\odot}}$ and $5.5-6.3\ \rm{M_{\odot}}$.
The tertiary star, also in the main sequence, is significantly more massive than both stars in the binary, with a mass of $14.5-16\ \rm{M_{\odot}}$.
The orbital configuration of the tertiary is more complex, with a mutual inclination of $i=16.8^{+4.2}_{-1.4}\ \rm{deg}$  and an eccentricity of $e=0.3$.
While the current configuration is dynamically stable, the origin and fate of \event\ is uncertain.

In this \textit{Letter}, we consider the progenitor scenario in which \event\ was originally a `2+2' quadruple-star system, with the now tertiary being originally similar to the companion binary.
In this scenario, an inclined orbit leads to ZLK oscillations \citep{1910AN....183..345V,1962AJ.....67..591K,1962P&SS....9..719L} that result in a contact phase for the more massive binary.
Assuming this contact phase leads to a stellar merger, the system reduces from an inclined `2+2' configuration into the triple configuration detected as \event.
This scenario circumvents the issues that arise if the three stars are assumed to be formed simultaneously.
Such a massive tertiary star, which has a shorter Kelvin-Helmholtz timescale, would have reached the main sequence earlier, disrupting the remaining natal material, and likely discontinuing the formation of the inner binary \citep{2022MNRAS.511.4710E}.
The (in)validation of this scenario will be fundamental to improving our understanding of stellar formation and reduction in multiple-star systems.

\section{Methods and Results}
Consider a first inner binary ($\A$) with masses $m_1$ and $m_2$,
and a distant second inner binary ($\B$) with masses $m_3$ and $m_4$, with an outer orbit ($\mathrm{out}$) around the centre of mass of the whole system.
For each orbit, we denote the semi-major axis $a_k$ (where $k$=A,B, or out),
eccentricity vector $\mathbf e_k=e_k \hat {\mathbf e}_k$, and angular momentum vector
$\mathbf L_k=L_k\hat{\mathbf L}_k=\mu_k[GM_k a_k (1-e_k^2)]^{1/2}\hat{\mathbf L}_k$, where $\mu_k$ and $M_k$ are reduced mass and total mass, and $\hat{\mathbf L}_k$ is the unit vector, respectively.
We thus define the inclination angles as $i_{\A(\B)}=\cos^{-1}(\hat{\mathbf L}_{\A(\B)}\cdot\hat{\mathbf L}_\OUT)$.

\textbf{Equations of motion.}
the orbital evolution of the quadruple system can be studied by expanding the Hamiltonian
and averaging over both the inner and outer orbits (double averaging).
For binary $A(B)$, the angular momentum and eccentricity
evolve according to
\begin{equation}\label{eq:FullKozai1}
    \frac{d \mathbf{L}_{A(B)}}{dt} =\frac{d \mathbf{L}_{A(B)}}{dt}\bigg|_{\zlk},
\end{equation}
\begin{equation}\label{eq:FullKozai2}
    \frac{d \mathbf{e}_{A(B)}}{dt} =\frac{d \mathbf{e}_{A(B)}}{dt}\bigg|_{\zlk}+\frac{d \mathbf{e}_{A(B)}}{dt}\bigg|_{\gr}+\frac{d\mathbf{e}_{A(B)}}{dt}\bigg|_{\tide}.
\end{equation}
    
Here, the ZLK terms refer to the oscillations of the inner binary
when the mutual inclination angle, $i_{\rm{A(B)}}$, is sufficiently high \citep[e.g.,][]{1910AN....183..345V,1962P&SS....9..719L,1962AJ.....67..591K}.
The associated ZLK timescale is
\be\label{eq:ZLKtimescale}
t_{\zlk,\A(\B)}=\frac{1}{n_{\A(\B)}}\frac{M_{\A(\B)}}{M_{\B(\A)}}\bigg(\frac{a_\OUT\sqrt{1-e^2_\OUT}}{a_{\A(\B)}}\bigg)^3,
\ee
where $n_{\A(\B)}=(G M_{\A(\B)}/a_{\A(\B)}^3)^{1/2}$ is the mean motion of binary $\A(\B)$.

For the outer orbit, the time evolution is given by 
\begin{equation}
\frac{d \mathbf{X}_\OUT}{dt}=\frac{d \mathbf{X}_\OUT}{dt}\bigg|_\A+\frac{d \mathbf{X}_\OUT}{dt}\bigg|_\B\label{eq:jout_eout}, \hfill \mathbf{X}=\{\mathbf{L},\mathbf{e}\}.  
\end{equation}
Note that the outer binary's angular momentum and eccentricity
are affected by Newtonian potential from both binaries $\A$ and $\B$.
The explicit equations of motion can be found in \citet{2019MNRAS.483.4060L}.

\textbf{Additional resonances in a `2+2' quadruple.}
Quadruple systems result in ZLK resonances that lead to eccentricity cycles in a wider range of inclinations than the triple-system configuration counterpart \citep{2015MNRAS.449.4221H,2016MNRAS.461.3964V,2017MNRAS.470.1657H,2018MNRAS.474.3547G}. This quadruple effect has been noted to enhance the merger rate of black hole binaries \citep[][]{2019MNRAS.483.4060L} and formation fraction of transiting giant planets \citep[][]{2021MNRAS.501..507O}. 
Quadruple systems may be in '3+1' or '2+2' configurations. Here, we discuss the ZLK eccentricity excitation as a mean of stellar reduction from a `2+2' configuration into a triple.

Considering the simple case where binary $\B$ and the outer binary stay circular ($e_\B=e_\OUT=0$), the angular momentum axis of the outer binary is driven into precession around the
$\mathbf{L}_{\B+\OUT}\equiv\mathbf{L}_\B+\mathbf{L}_\OUT$ axis following
\be\label{eq:j2vecB}
\frac{d{\hat{\mathbf{L}}_\OUT}}{dt}\bigg|_\B
=\Omega_\OUT \hat{\mathbf{L}}_{\OUT}\times\hat{\mathbf{L}}_{\B+\OUT}
\simeq\bigg(\frac{3}{4}\frac{\cos i_\B}{t_{\zlk,\B}}\bigg)\hat{\mathbf{L}}_{\OUT}\times\hat{\mathbf{L}}_{\B+\OUT}.
\ee
Since the outer binary drives ZLK oscillations of binary $\A$ on a timescale $t_{\zlk,\A}$,
we can define the dimensionless parameter
\be\label{eq:beta}
\beta\equiv\Omega_\OUT t_{\zlk,\A}=\frac{3}{4}\cos i_\B\bigg(\frac{a_\B}{a_\A}\bigg)^{3/2}
\bigg(\frac{m_{\A}}{m_{\B}}\bigg)^{3/2}.
\ee
The qualitative nature of the dynamics is determined by the value of $\beta$. 
When $\beta\ll1$, binary $\B$ essentially behaves like a single mass ($m_{\B}$),
and the standard ZLK problem of a `2+1' hierarchical triple applies.
When $\beta\gg1$, the problem reduces to that of canonical ZLK oscillations, with $\hat{\mathbf{L}}_{\B+\OUT}$ serving
as the effective $\hat{\mathbf{L}}_\OUT$. 
Finally, when $\beta\sim 1$, a secular resonance occurs that generates
large $i_{\A(\B)}$ even for initially low-inclination systems, and this resonantly excited inclination then leads to ZLK oscillations of the inner binary.

\textbf{The effect of short-range effects on eccentricity cycles.}
When the pericentre separation $a_{\rm{p}}=a(1-e)$ is small, short-range effects (SRE) from, e.g., General Relativity (GR) and tides, can become significant. 
We implement these additional effects in our model following \citet{2015MNRAS.447..747L} as
\be
\frac{d \mathbf{e}_\A}{dt}\bigg|_{\rm SRE}
=\frac{\varepsilon_{\rm SRE}}{t_{\zlk,\A}}f_{\rm SRE}(e_A)\hat{\mathbf L}_\A\times\mathbf e_\A,
\ee
where $\rm{SRE}=\{\gr,\tide \}$, $f_\gr(e)=1/(1-e^2)$ and $f_\tide(e)= (1+3e^2/2 + e^4/8)/(1-e^2)^5$, and
\begin{align}
\varepsilon_\gr&=\frac{3GM_{\A}^2a_\OUT^3(1-e^2_\OUT)^{3/2}}{c^2a_\A^4M_{\B}}\label{eq:epsilonGR},\\
\varepsilon_\tide & =\frac{15M_{\A}a_\OUT^3(1-e^2_\OUT)^{3/2}(m_1^2k_2R_2^5+m_2^2k_1R_1^5)}{a_\A^8m_1m_2M_{\B}}\label{eq:epsilonTIDE}
\end{align}
with $k_1=k_2=0.014$ being the apsidal motion constant of a main sequence star,
$R_1$ and $R_2$ being the radius of $m_1$ and $m_2$.
The coefficients $\varepsilon_\gr$ and $\varepsilon_\tide$ represent the strength of the short-range effects
\citep[e.g.,][]{2007ApJ...669.1298F}.
Note that we consider the tidal distortions of both stars in Equation (\ref{eq:epsilonTIDE}).
Short-range effects can limit the growth of eccentricity and even suppress standard ZLK oscillations.
In the quadrupole approximation, the maximum eccentricity for all values of mutual inclinations, called $e_\li$, can be calculated analytically \citep[e.g.,][]{2015MNRAS.447..747L,2017MNRAS.467.3066A} using energy and angular momentum conservation leading to
\ba\label{eq:ELIM}
&&\frac{3}{8}(j_{\A,\li}^2-1)\left[-3+\frac{\eta_\A^2}{4}\left(\frac{4}{5}j_{\A,\li}^2-1\right)\right]+
\varepsilon_\gr \left(1-\frac{1}{j_{\A,\li}}\right)\nonumber\\
&&+\frac{\varepsilon_\tide}{15} \left(1-\frac{1+3e^2_{\A,\li}+3e^4_{\A,\li}/8}{j_{\A,\li}^9}\right)=0,
\ea
where $j_{\A,\li}\equiv\sqrt{1-e_{\A,\li}^2}$ and
$\eta_\A\equiv(L_\A/L_\OUT)_{e_\A=0}$.

Similarly, the limiting eccentricity of the binary $\B$ ($e_{\B,\li}$) can be obtained as well
by replacing the relevant parameters in Equation (\ref{eq:ELIM}).
Note that the analytic expression for $e_\li$ 
was derived for the standard `2+1' ZLK problem with short-range effects,
but it remains valid when the octupole effect is included \citep[]{2015MNRAS.447..747L,2017MNRAS.467.3066A},
and it is also valid for the `2+2' problem \citep[e.g.,][]{2019MNRAS.483.4060L,2021MNRAS.501..507O}.

\textbf{Initial conditions.} 
For simplicity, we will consider the scenario in which each binary has equal-mass stellar components.
We follow \cite{2000MNRAS.315..543H} to determine the radii for these stars at metallicity close to Solar ($Z=0.0142$).
We consider a `6+6' binary where each star has a mass of $6\ \rm{M_{\odot}}$ and a ZAMS radius of $R_{\rm{6,ZAMS}}=2.8\ \rm{R_{\odot}}$, similar to the binary from \event.
The initial eccentricity and semi-major axis are $e_{6+6}=0.001$ and $a_{6+6}=10.32\ \rm{R_{\odot}}$, respectively, which results in an orbital period of $\approx 1.10$~d, and inclination $i_{6+6}=17$ deg.
For the tertiary progenitor, we consider an `8+8' binary constituted of two $8\ \rm{M_{\odot}}$ stars with a ZAMS radius of $R_{\rm{8,ZAMS}}=3.3\ \rm{R_{\odot}}$ each.
We assume the `8+8' binary is also initially almost circular ($e_{6+6}=0.001$), and explore the inclination ($i_{8+8}$) and separation ($a_{8+8}$) parameter space.

We sample the initial inclination uniformly in the range $-1 \leq \cos{(i_{8+8}/\rm{deg})} \leq 1$.
For the separation of the `8+8' binary, we consider the parameter space that satisfies two conditions.
The first one is that the orbit of the system is large enough that the component stars are not filling their Roche lobe at ZAMS.
We follow \cite{1983ApJ...268..368E} to estimate the Roche-lobe radius ($R_{\rm{RL}}$) of an equal mass binary as $R_{\rm{RL}}=a_{\rm{p}} f(q=1) \approx 0.38 a_{\rm{p}}$, where $q$ is the mass ratio and $a_{\rm{p}}=a(1-e)$ is the periastron.
The second condition is that the system is dynamically stable, following the \cite{2001MNRAS.321..398M} criterion
\be\label{eq:stability}
\frac{a_\OUT}{a_\A}>2.8\bigg(1+\frac{m_{\B}}{m_{\A}}\bigg)^{2/5}\frac{(1+e_\OUT)^{2/5}}{(1-e_\OUT)^{6/5}}\bigg(1-\frac{0.3 i_\A}{180\ \rm{deg}}\bigg).
\ee
If these two criteria are satisfied, we simulate the system.
For each simulation, we determine if the dynamics lead to a merger.
We consider a stellar merger occurs if the radius of the star reaches the second Lagrangian point, $L_2\approx 1.32 R_{\rm{RL}}$ \citep{2016A&A...588A..50M}.

Finally, the outer orbit and less massive `6+6' binary are always initialised in the same way in order to be consistent with \event.
For the outer orbit, $a_{\rm{out}}=176.33\ \rm{R_{\odot}}$ and $e_{\rm{out}}=0.3$, which result in an outer orbital period of $\approx 51$ d for the `6+6' and `8+8' mass configuration.

\textbf{Detailed evolution of a quadruple.} 
We present the orbital evolution of the binary elements of a quadruple in an orbital configuration.
The orbit of the more massive `8+8' binary is initialised with an inclination of $i_{8+8}=75$ deg and separation of $a_{8+8}=19.4\ \rm{R_{\odot}}$. For this system, the ZLK timescale ($\tau_{\rm{ZLK,8+8}} \sim$ yr) is significantly shorter than the thermal and nuclear timescales; therefore, we neglect the role of stellar evolution. We follow the system for $500\times \tau_{\rm{ZLK,8+8}}$ to make sure no dynamical instabilities arise.

\begin{figure}
	\includegraphics[trim=0.75cm 7.5cm 1cm 7.5cm,clip,width=\columnwidth]{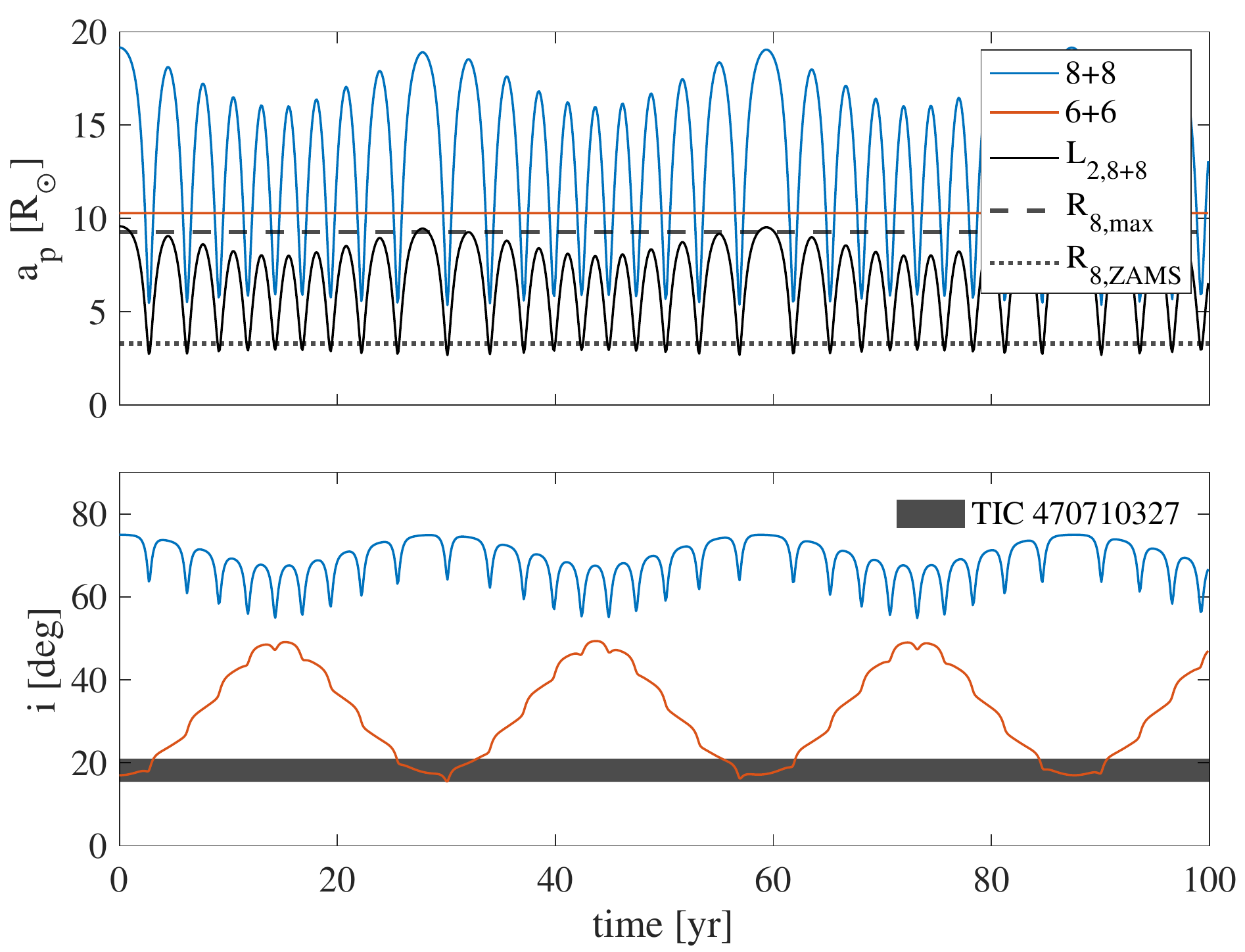}
    \caption{
    Dynamical evolution of a quadruple leading to a \event-like triple.
    The more (`8+8') and less (`6+6') massive binaries are shown in blue and red, respectively. 
    The initial separation and inclination are $a_{8+8}=17.2\ \rm{R_{\odot}}$ and $i_{8+8}=75$ deg, respectively.
    Top panel: periastron as a function of time. 
    The solid black line represents the instantaneous location of the $L_2$ point for the `8+8' binary.  
    The dotted and dashed black lines indicate the ZAMS ($R_{\rm{8,ZAMS}}$) and maximum main-sequence radius ($R_{\rm{8,max}}=9.3\ \rm{R_{\odot}}$) of the $8\ \rm{M_{\odot}}$ star, respectively.
    Bottom panel: inclination as a function of time. The grey shaded region corresponds to the inferred inclination values of the mutual inclination, between the inner and outer orbit, of \event.
    }
    \label{fig:evolution}
\end{figure}

We zoom in the first $100$~yr of the simulation to follow the evolution of the separation of each binary, as well as their inclinations with respect to the outer angular momentum vector (Figure \ref{fig:evolution}).
The inclination oscillates between $15.5 \leq i_{6+6}/\rm{deg} \leq 49.2$ and $54.9 \leq i_{8+8}/\rm{deg} \leq 75.0$, with a periodicity of $\approx 30$~yr.
This configuration results in an early contact phase for the `8+8' binary, a few years\footnote{Other configurations might delay the short-timescale merger several years or decades.} after the beginning of the simulation (ZAMS).
We consider this quadruple as a progenitor candidate of \event.

\begin{figure}
	\includegraphics[trim=0.75cm 7.5cm 1cm 7.5cm,clip,width=\columnwidth]{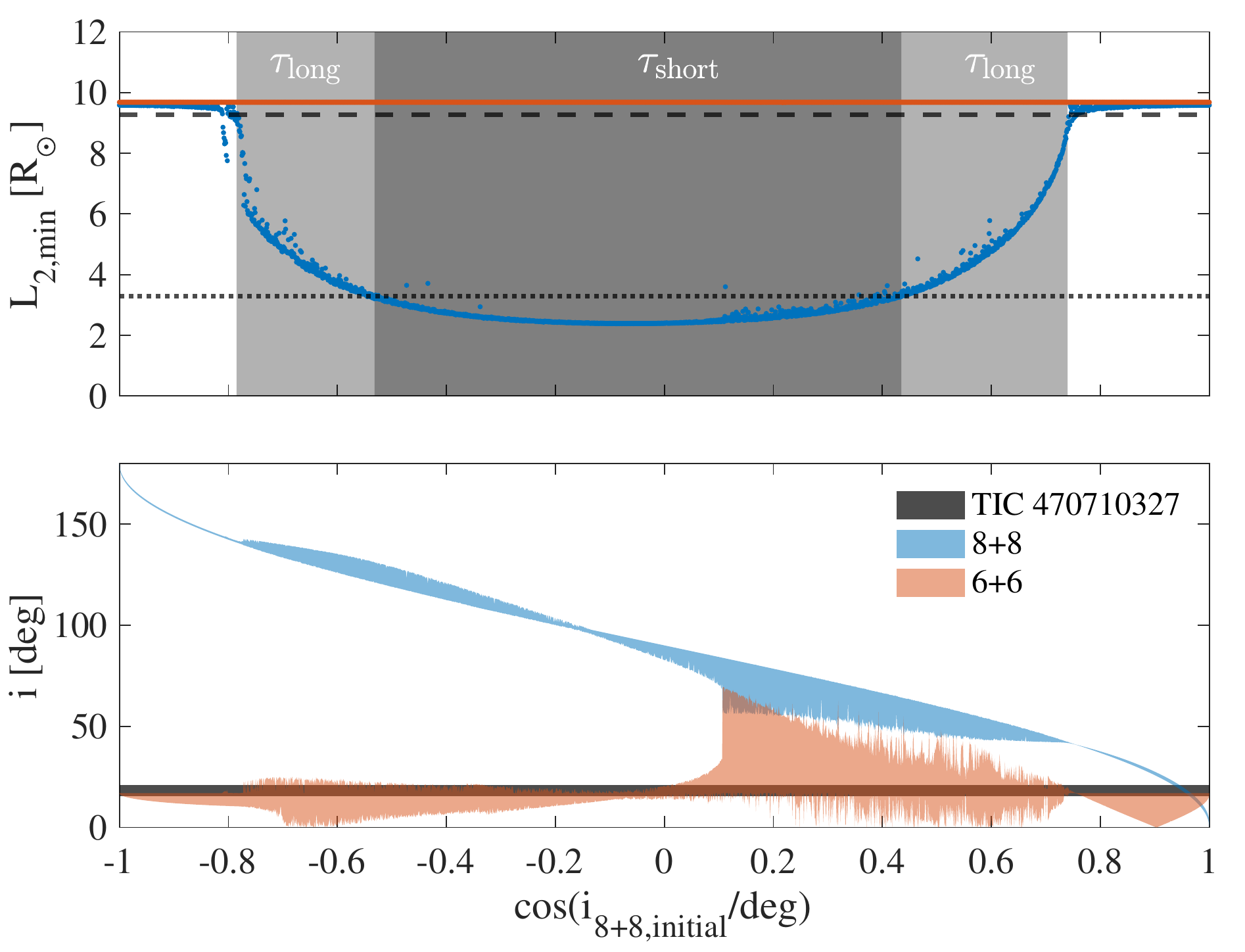}    \caption{
    Inclination phase space of a quadruple with initial separation of $a_{8+8}=19.4\ \rm{R_{\odot}}$.
     The abscissa shows the cosine of the initial inclination.
    The lines and colour are the same as in Figure \ref{fig:evolution}.
    Top panel: Minimum $L_2$ as a function of initial inclination.
    The regions are divided in those which we expect lead to a merger on a short timescale ($\tau_{\rm{short}}$ in light grey) or alternatively on a longer timescale ($\tau_{\rm{long}}$ in grey), the latter driven by radial expansion.
    Bottom panel: the range of inclinations as a function of the initial inclination.
    }
    \label{fig:phase_space}
\end{figure}

\textbf{Phase space of quadruples leading to a contact phase.} 
In order to explore the full inclination and separation parameter space that can reduce `2+2' systems into triples, we simulate $\sim 10^5$ quadruples.
For each allowed separation of the massive `8+8' binary, we retrieve the maximum eccentricity in order to estimate the minimum periastron distance ($a_{\rm{p,min}}$) which is then compared to $L_{2,8+8}$, our threshold for a contact phase.
In Figure \ref{fig:phase_space} we show the results of this analysis for a representative quadruple with $a_{8+8}=19.4\ \rm{R_{\odot}}$.
For systems with initially small ($i_{\rm{8+8,initial}}\lesssim 40$ deg) or large ($i_{\rm{8+8,initial}}\gtrsim 140$ deg) inclinations, the separation and eccentricity of the `8+8' binary remain effectively unchanged, even when both binaries are modulating their inclinations (Figure \ref{fig:phase_space}).
In between the aforementioned inclinations, there is a region where the eccentricity enhancement does lead to a decrease in the periastron distance: that is the ZLK window.
Deep within the ZLK window there is a \textit{contact} window where $R_{\rm{8,ZAMS}}>L_{2,8+8}$: this is the region of \event-like progenitors that occur a short timescale ($\tau_{\rm{short}}\sim 10-1000$ yr).
There is another region where contact won't occur at ZAMS, but might occur later in the evolution of the system, when stellar evolution leads to radial expansion on a longer timescale ($\tau_{\rm{long}}\gg 1000$ yr).
We consider these \textit{short}- and \textit{long}- timescale contact windows as regions of interest for mergers.
The short window is between the limit angles $i_{8+8}^-\leqslant i_{8+8}\leqslant i_{8+8}^+$.
If we assume the orbital orientation of the first inner binary is distributed isotropically, 
the fraction of induced contact binaries is given by $f=|\cos i_{8+8}^+-\cos i_{8+8}^-|/2$.
We calculate the long window similarly.

\begin{figure}
	\includegraphics[trim=1.0cm 7.5cm 1.0cm 7.5cm,clip,width=\columnwidth]{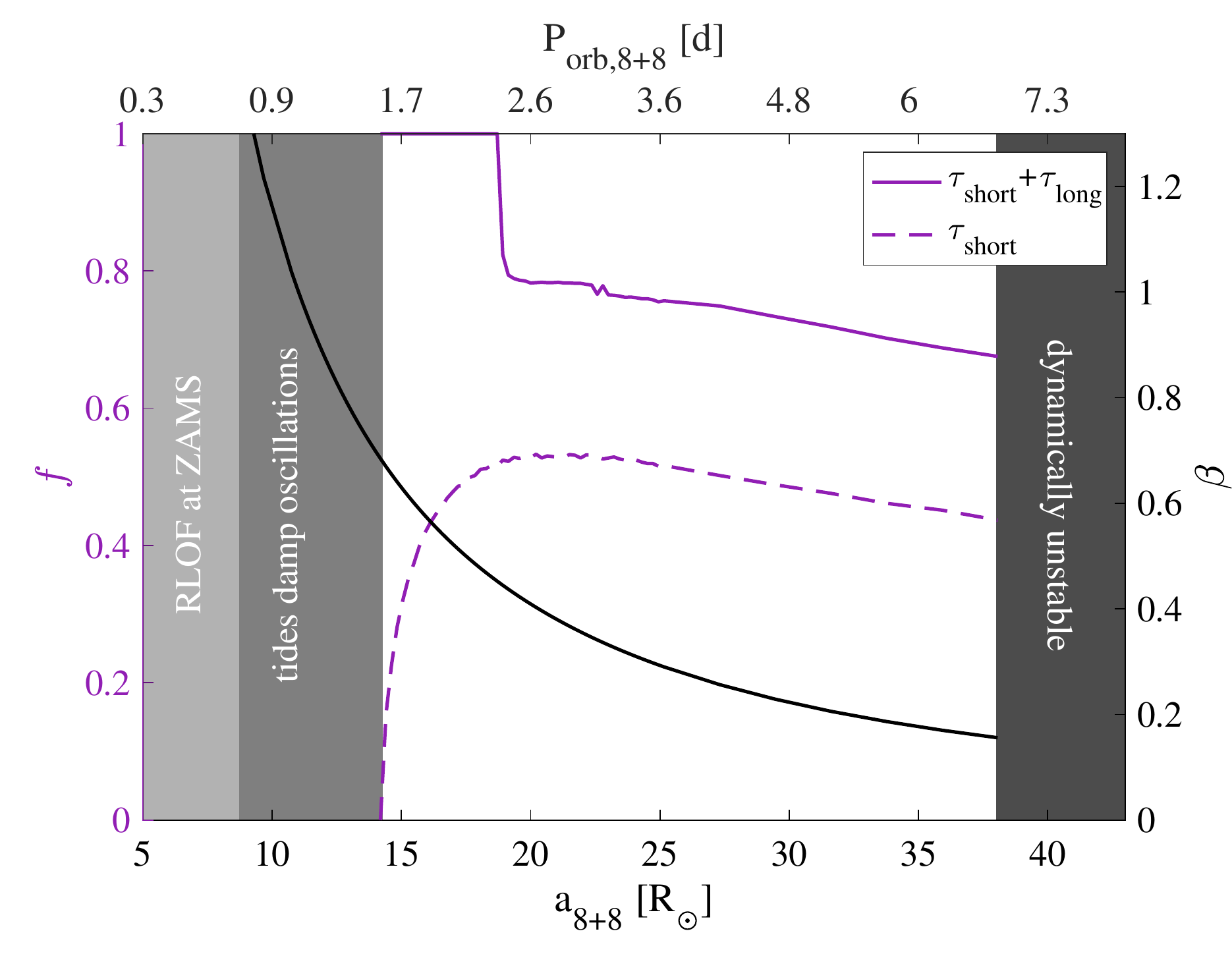}
    \caption{
    Summary of the initial separation parameter space of the quadruples simulated.
    The abscissa shows both the initial semi-major axis (bottom) and orbital period (top) of the more massive `8+8' binary. 
    The ordinate shows the contact fraction $f$ (left, purple) and the $\beta$ factor of the simulation (right, black) of `2+2' quadruples where the `8+8' binary results in a merger.
    The solid purple line includes the merger fraction from both the short and long contact window; the dashed purple line includes only the fraction of the short contact window (see Figure \ref{fig:phase_space} for a visual representation of these contact windows).
    For close ($\lesssim 19\ \rm{R_{\odot}}$) binaries, radial expansion will lead to a contact phase later in the main sequence.
    }
    \label{fig:check}
\end{figure}

In Figure \ref{fig:check} we show this fraction as a function of initial separation.
\event-like progenitors of an `8+8' binary are restricted between $14 \lesssim a_{8+8}/{\rm R_{\odot}} \lesssim 38$, which corresponds roughly to $1 \lesssim P_{\rm orb,8+8}/{\rm d} \lesssim 7$.
Short period ($\lesssim 1.6$ d) binaries would either be in contact at ZAMS or dominated by short-range effects, which would suppress eccentricity from potential ZLK oscillations (Equation \ref{eq:ELIM}).
Long period ($\gtrsim 6.8$ d) binaries do not satisfy the stability criteria (Equation \ref{eq:stability}).
Marginally stable binaries ($5 \lesssim P/\rm{d} \lesssim 6.8$) are prone to non-secular fluctuations that are not accounted for in the averaging procedure and could further increase the maximal eccentricity \citep{2016MNRAS.458.3060L,2018MNRAS.481.4907G}.
At all separations, the merger fraction is larger if we consider the radial expansion associated to the long-timescale contact window. 
The schematic of Figure \ref{fig:check} is representative of the `8+8' configuration, but should not vary much within the $14.5-16\ \rm{M_{\odot}}$ mass constrain for the tertiary star of \event.
We also performed simulations with a `10+6' binary and arrive to similar results.

\section{Discussion and Conclusions}

\textbf{$\tau$ Sco and massive mergers.} 
$\tau$ Sco is a bright magnetic massive B0.2V blue-straggler star in the upper Sco association \citep[][and references therein]{2006MNRAS.370..629D}: it is also slowly rotating, possess nitrogen excess, and it is magnetised \citep[e.g.,][and references therein]{2021MNRAS.504.2474K}. 
\cite{2006MNRAS.370..629D} presented spectropolarimetery of $\tau$ Sco, and inferred a medium-strength ($\sim 0.5$ kG) magnetic field with a complex, non-dipolar structure \citep[see also][]{2016A&A...586A..30K}.
\cite{2021MNRAS.504.2474K} studied the effect of rotation and magnetic fields in massive stars in the context of $\tau$ Sco.
They find that nitrogen chemical enrichment challenges the single-star scenario, and suggest this might hint toward a binary origin.
While non-axisymmetric magnetic equilibrium in stars has been explored in the literature \citep{2008MNRAS.386.1947B}, the magnetic field evolution of such configurations remains to be fully understood.

\cite{2019Natur.574..211S} explored the magnetic-field amplification during the merger of massive stars in the context of $\tau$ Sco.
They used three-dimensional magnetohydrodynamical methods to model the merger of a 9-Myr-old core-hydrogen-burning `9+8' binary at $Z=0.0142$.
The turbulent merger amplifies the ($\sim \mu \rm{G}$) magnetic field exponentially and results in a $\approx 16.9\ \rm{M_{\odot}}$ remnant with $\lesssim 0.1\ \rm{M_{\odot}}$ in a disk around it.
The simulation ends $\approx$10 d after the merger, with a remnant that has a surface magnetic-field strength of $\approx 9\ $ kG.
At this point, the magnetic field of the merger model is assumed to be dipolar, and the merger remnant is evolved using a one-dimensional stellar evolution code.
During the thermal relaxation phase, the merger product becomes over-luminous and rapidly spinning, getting close to critical rotation and shedding mass.
This mass loss removes some ($\approx 7\%$) of the angular momentum of the remnant, which then decreases in luminosity by an order of magnitude; internal restructuring of the star leads to a slowly-rotating $\approx 16.9\ \rm{M_{\odot}}$ main-sequence star with a strong ($\approx 9\ $ kG), long-lived ($\sim 10^7$ yr) surface magnetic field.
The merger product is likely to be a rejuvenated blue star \citep[e.g.,][and references therein]{2020MNRAS.495.2796S}.
Recently, \cite{2022NatAs.tmp...35W} proposed stellar mergers as the origin of the blue main-sequence band in young star clusters.
One of the signatures of such stars, besides their colour, is their low fractional rotational velocities ($v_{\rm{rot}}/v_{\rm{crit}}\approx 0.35$), which could be the result of magnetic braking.
Slow rotation and high magnetic fields in the tertiary star of \event would support the proposed scenario.
Moreover, the detection of a magnetic field would be of paramount importance to the study and understanding of magnetism in massive stars.

\textbf{\event\ as the host of an early massive merger.}
We have demonstrated that ZLK oscillations can prompt a contact phase in a`2+2' quadruple and potentially reduce it into a triple-star configuration.
For \event, the massive tertiary could be a stellar merger remnant if the inclination around ZAMS was within the short-timescale contact window (Figure \ref{fig:phase_space}).
We assume that is the case, and that our `8+8' stellar merger might evolve similarly to the `9+8'merger from \cite{2019Natur.574..211S}.
That implies that the tertiary of \event\ should be slowly rotating and could have a magnetic field as large as $\approx 9\ $ kG.

We make a simple estimate of the magnetic field evolution by considering a magnetic dipole configuration and that the magnetic flux is frozen into the plasma \citep{1942Natur.150..405A}.
Conservation of magnetic flux results in $B_{\rm{p,min}}=B_{\rm{p,ZAMS}}(R_{\rm{ZAMS}}/R_{\rm{max}})^2$,
where $B_{\rm{p}}$ is the polar magnetic field strength, during the main sequence, at the photosphere.
For a $16.0\ \rm{M_{\odot}}$ star at $Z=0.0142$, $R_{\rm{ZAMS}}=5.0\ \rm{R_{\odot}}$ and $R_{\rm{max}}=15.2\ \rm{R_{\odot}}$ \citep{2000MNRAS.315..543H}.
The magnetic field could decrease from $B_{\rm{p,ZAMS}}\approx 9.0$ kG to $B_{\rm{p,max}}\approx 1.0$ kG within the main sequence.
We consider these values as proxy limits of the magnetic field of the tertiary of \event.
However, a dipolar magnetic field is a simplified assumption with respect to the magnetic-field structure of an astronomical stellar merger.

\cite{2019MNRAS.485.5843K} predicts that rotating, magnetised massive stars might eventually lead to notable surface enrichment of nitrogen.
However, our short-timescale mergers would occur very early in the evolution of the system, and are predicted to become slowly rotating shortly ($\sim 1000$ yr) after the merger \citep{2019Natur.574..211S}.
However, the mergers on the long window might possess some chemical anomalies.

\textbf{Stellar reduction.}
The scenario presented here proposes a `2+2' quadruple origin for a triple-star system.
This scenario highlights the role of stellar reduction in massive stars, and suggests that stellar multiplicity likely decreases as a function of time.
While most massive stars ($\gtrsim 5\ \rm{M_{\odot}}$) are observed in binary or multiple-star configurations \citep{2017ApJS..230...15M}, this multiplicity might be even higher at early stages of their evolution; i.e., some singles were initially binaries, some binaries were initially triples, and so on.
A `2+2' quadruple origin for \event\ could suggest that the massive tertiary might have formed in a very similar way to the less massive, close binary \citep[see][for a similar formation scenario for low-mass triples]{2018AJ....155..160T}.
We have shown that the orbital configurations can change in short and long timescales \citep[see also, e.g.,][]{2001ApJ...562.1012E}.
Understanding the role of dynamics in (proto-)stellar evolution will help us elucidate the true initial distributions of multiple-star systems.

This reduction scenario has implications on the orbital configuration of massive, multiple-star systems.
Compact ($\sim$ d) binaries in inclined orbits, i.e., those within the short window, are unlikely to be long-lived stable astronomical configurations, given that they would have merged around ZAMS, reducing the stellar system to one without significant inclination.
This is consistent with populations of low-mass hierarchical triples \citep[e.g.,][]{2017ApJ...844..103T,2022MNRAS.tmp..922R}.
We suggest a deficiency of highly-inclined systems as a smoking-gun signature for populations of compact binaries in multiple-star systems.
Population synthesis predicts that $\approx$10\% of all stars experience a merger with a companion \citep[e.g.,][]{1992ApJ...391..246P}, a fraction similar to that ($\approx$7 \%) of magnetic B- and O- type stars \citep[e.g.,][]{2015A&A...582A..45F,2017MNRAS.465.2432G}.
These stars also have a higher-multiplicity fraction, which likely increases the probability of mergers.
We suggest exploring the role of stellar formation and multiplicity in early mergers of massive stars in the context of magnetic stars \citep[see also][]{2022NatAs.tmp...35W}.

We briefly comment on the possibility that the merger of the tertiary occurred from a stellar collision with the triple system. 
The properties of the stellar merger remnant in the collision scenario should be similar to the ones discussed here; however, there should be kinematic differences, as the collision will result in additional mass loss \citep{2013MNRAS.434.3497G} and will likely modify the velocity and trajectory of the system.
Alternatively, the triple could have formed via dynamical capture.

Finally, we highlight
\cite{2021ApJ...907L..19V}, which presented the role of compact binaries in triples, particularly as progenitors of sequential binary black-hole mergers.
Particularly, chemically homogeneously evolving binaries can only occur in within a narrow period range between $0.7 \lesssim P_{\rm{orb}}/\rm{d} \lesssim 4$ \citep[e.g.,][and references therein]{2020MNRAS.499.5941D}.
While chemically homogeneously evolving binaries are predicted to be significantly more massive than our `8+8' binary, and therefore they are not the likely progenitors of \event, they could be in similar configurations that experience dynamical effects and stellar reduction during the main sequence (cf. Figure \ref{fig:check}).

\textbf{Conclusions.}
Here we propose a quadruple origin for the massive, compact hierarchical triple \event.
Our proposed `2+2' orbital configuration results in ZLK oscillations that prompt the merger of the more massive binary, leading to the triple configuration we detect today.
Such stellar merger is generally predicted to result in a highly-magnetised ($\sim 1-10$ kG) slowly-rotating blue star on the main sequence.
This formation scenario predicts that highly-inclined triple- and quadruple- star systems will experience stellar mergers and reduce to co-planar binary- and triple- star systems, respectively.
Disentangling the nature of such peculiar systems is fundamental for progress in the emerging field of multiple massive-stellar formation and evolution.

\section*{Acknowledgements}
We thank Warrick Ball, Alexey Bobrick, Pablo Marchant, Max Moe, and Martin Pessah for useful discussions.
A.V-G. received support through Villum Fonden grant no. 29466.
B.L. gratefully acknowledges support from the European Union’s Horizon 2020 research and innovation program under the Marie Sklodowska-Curie grant agreement No. 847523 ‘INTERACTIONS’. 
D.R.A-D. is supported by the Stavros Niarchos Foundation (SNF) and the Hellenic Foundation for Research and Innovation (H.F.R.I.) under the 2nd Call of ``Science and Society’' Action Always strive for excellence – ``Theodoros Papazoglou’' (Project Number: 01431). 
E.R-R. thanks the Heising-Simons Foundation and the NSF (AST-1911206, AST-1852393, and AST-1615881) for support.
M.S.F. gratefully acknowledges support provided by NASA through Hubble Fellowship grant HST-HF2-51493.001-A awarded by the Space Telescope Science Institute, which is operated by the Association of Universities for Research in Astronomy, In., for NASA, under the contract NAS 5-26555. 

\section*{Data Availability}
The data underlying this article will be shared upon reasonable request to the corresponding author.



\bibliographystyle{mnras}
\bibliography{references} 


\bsp	
\label{lastpage}
\end{document}